# Controlling Delay-induced Hopf bifurcation in Internet congestion control system

Dawei Ding[a,b,*], Jie Zhu[a], Xiaoshu Luo[c], Yuliang Liu[a]

a. Department of Electronic engineering, Shanghai Jiao Tong University, Shanghai 200240, China

b. School of Electronic science and technology, Anhui University, Hefei 230039, China

c. College of Physics and Electronic Engineering, Guangxi Normal University, Guilin 541004, China

**Abstract**

This paper focuses on Hopf bifurcation control in a dual model of Internet congestion control algorithms which is modeled as a delay differential equation (DDE). By choosing communication delay as a bifurcation parameter, it has been demonstrated that the system loses stability and a Hopf bifurcation occurs when communication delay passes through a critical value. Therefore, a time-delayed feedback control method is applied to the system for delaying the onset of undesirable Hopf bifurcation. Theoretical analysis and numerical simulations confirm that the delayed feedback controller is efficient in controlling Hopf bifurcation in Internet congestion control system. Moreover, the direction of the Hopf bifurcation and the stability of the bifurcating periodic solutions are determinated by applying the center manifold theorem and the normal form theory.

## 1. Introduction

Recently, Internet congestion control algorithms have been the focus of intense study since the seminal work of Kelly [1]. These congestion control algorithms can be divided into three classes: primal algorithms, dual algorithms and primal-dual algorithms [2]. In primal algorithms, the users adapt the source rates dynamically based on the route prices (the congestion signal generated by the link), and the links select a static law to determine the link prices directly from the arrival rates at the links. While in dual algorithms, the links adapt the link prices dynamically based on the link rates, and the users select a static law to determine the source rates directly from the route prices and the source parameters. Primal-dual algorithms combine these two schemes and dynamically compute both user rates and link prices.

[*] Corresponding author. Tel. +86 21 34205433   Fax. +86 21 34205432
Email: dwding@sjtu.edu.cn, dwdingsjtu@gmail.com

The research of congestion control system in the Internet can be divided into two categories. On one hand, many researchers have paid their attentions to stability properties of the congestion control system. In [3], authors have studied the local asymptotic stability under communication delay of the primal algorithm. Later Massoulie has extended these local stability results to general network topologies and heterogeneous delays [4]. In [5], the authors have analyzed the global, asymptotic stability and semiglobal exponential stability properties of congestion control scheme under fixed heterogeneous delays, and for general network topologies. In [6], authors have extended the framework of [1] and given a condition for its local stability under delay using the generalized Nyquist criterion. Sichitiu and Bauer have studied the asymptotic stability of congestion control systems with multiple sources [7]. Tian has proved the local asymptotic stability of congestion control with diverse communication delays in [8, 9]. Then he also derived the sufficient conditions for local asymptotic stability of the second-order congestion control algorithm [10]. Ranjan et.al have analyzed Kelly's optimization framework for a rate allocation problem and provide stability conditions with arbitrary fixed communication delays [11]. In [12], Wang and Eun have studied the local and global stability of TCP-newReno/RED under many flows.

On the other hand, also other researchers are interesting in what the system will be when the congestion control system loses its stability and focus on its dynamic behavior such as bifurcation and chaos. One of the most cited paper about the chaotic nature of TCP is [13]. Using some discrete-time models, researchers have shown that TCP/RED systems become chaotic dynamics with variability in RED parameters [14-17]. In [18-20], the authors have shown that the existence of Hopf bifurcation in one-order and two-order REM with a single link and single source. Guo et al. have investigated the Hopf bifurcation of the exponential RED algorithm with communication delay [21, 22]. Raina has studied the local bifurcation of the fair dual with proportional and TCP fairness, and the delay dual algorithms by choosing a non-dimensional parameter as bifurcation parameter [23]. In [24], with communication delay as bifurcation parameter, we have studied the Hopf bifurcation in a fluid-flow model of Internet congestion control system.

In reality, the complex dynamic behavior means that the system changes from a stable state to an unstable one, which is sometimes harmful to the system. Therefore, researchers have attempted a lot of methods to delay or even avoid this kind of behaviors. In [25], authors have proposed a Time-Delayed Feedback Controller (TDFC) to delay the period-doubling bifurcation and increase the stability of TCP-RED system described in [14]. We have also applied a hybrid control strategy to control the bifurcation and chaos in a stroboscopic model of Internet congestion control system [26]. Chen and Yu have proposed a time-delayed feedback controller using polynomial function to delay the onset of Hopf bifurcation in an Internet congestion control model with a single router and single source [27]. Then in [28], Xiao and Cao have proposed a more general controller than that in [27] on the preservation of equilibrium points of the system. In [29], the TDFC method is combined with the primal algorithm for eliminating oscillation in the Internet.

In this paper, we deal with Hopf bifurcation control in a fair dual algorithm of congestion control system using a time-delayed feedback controller which is following the ideal of Pyragas [30]. Unlike the control issues in [26-28], we choose communication delay as bifurcation parameter. Since delay is the round-trip time of an indication signal between the user and the source, it may vary depending on the network congestion status. Meanwhile, the formulas for determining the direction of the Hopf bifurcation and the stability of bifurcating periodic solutions of controlled system are obtained by applying the center manifold theorem and the normal form theory. Finally numerical simulations are given to justify the validity of delayed feedback controller in bifurcation control of the Internet congestion control system.

The rest of this paper is organized as follows. In the next section, we study the existence of the Hopf bifurcation in the dual model of Internet congestion control system with or without control. In Section 3, based on the center manifold theorem and the normal form theory, the formulas for determining the stability of bifurcating solutions and the direction of the Hopf bifurcation are deduced. Then, numerical examples are given to verify the theoretic analysis. Finally, conclusion remarks are given in Section 5.

## 2. Existence of Hopf bifurcation of system with control

The dynamical representation of a dual congestion control system is as follows:

$$\dot{p}(t) = kp(t)(x(t-\tau) - c) \tag{1}$$

where $x(t) = f(p(t))$ is a nonnegative continuous, strictly decreasing demand function and has at least third-order continuous derivatives. The scalar $c$ is the capacity of the bottleneck link and the variable $p$ is the price at the link. While $k$ is a gain parameter.

We add a time-delayed force $h(p(t) - p(t-\tau))$ to the model (1), and then get the following controlled system:

$$\dot{p}(t) = kp(t)(x(t-\tau) - c) + h(p(t) - p(t-\tau)) \tag{2}$$

where the feedback gain $h$ is negative real number. It is obvious that the controlled system has the same equilibrium point as the original model (1).

The initial condition of Eq. (2) is specified by a real-valued continuous function

$$p(s) = \phi(s), s \in [-\tau, 0]$$

Let $p^*$ be an equilibrium point of (2). Then $p^*$ satisfy

$$x(p^*) = c \tag{3}$$

Define $u(t) = p(t) - p^*$ and take a Taylor expansion of Eq. (2) including the linear, quadratic and cubic terms, we obtain

$$\dot{u}(t) = hu(t) + b_2 u(t-\tau) + b_4 u(t)u(t-\tau) + b_5 u^2(t-\tau) + b_8 u(t)u^2(t-\tau) \\ + b_9 u^3(t-\tau) + O(u^4) \tag{4}$$

where

$$b_2 = kp^* x'(p^*) - h, b_4 = \frac{1}{2}kx'(p^*), b_5 = \frac{1}{2}kp^* x''(p^*),$$

$$b_8 = \frac{1}{6}kx''(p^*), b_9 = \frac{1}{6}kp^* x'''(p^*)$$

The linearized equation of Eq.(4) is

$$\dot{u}(t) = hu(t) + b_2 u(t-\tau) \tag{5}$$

and its characteristic equation is:

$$\lambda - h - b_2 e^{-\lambda \tau} = 0 \tag{6}$$

First, we examine when Eq.(6) has pure imaginary roots, i.e., $\lambda = \pm i\omega, \omega > 0$. Then separating the real and imaginary parts of (6), we get

$$\begin{cases} h + b_2 \cos \omega \tau = 0 \\ \omega + b_2 \sin \omega \tau = 0 \end{cases} \quad (7)$$

Therefore we obtain

$$\omega_0 = \sqrt{b_2^2 - h^2} \quad (8)$$

and

$$\tau_0 = \frac{1}{\omega_0} \arccos\left(-\frac{h}{b_2}\right) \quad (9)$$

Let $b = kp^* x'(p^*)$. From Eq.(8), we find that

$$b_2^2 - h^2 \geq 0$$

And considering that $h$ is negative, we have

$$0 \geq h \geq \frac{b}{2} \quad (10)$$

Next, we want to see whether (6) has roots with positive real parts when $\tau = \tau_0$. Let $\alpha + i\omega$ be a root of (6) with $\alpha$ and $\omega$ positive, then

$$\begin{cases} \alpha - h - b_2 e^{-\alpha \tau} \cos \omega \tau = 0 \\ \omega + b_2 e^{-\alpha \tau} \sin \omega \tau = 0 \end{cases} \quad (11)$$

Since $b = kp^* x'(p^*) < 0$ and considering (10), we get

$$b_2 = b - h \leq \frac{b}{2} < 0 \quad (12)$$

From the first equation of (11), we know

$$\frac{(2n+1)\pi}{2} < \omega \tau < \frac{(2n+3)\pi}{2}, \quad n = 0, 2, 4, \cdots \quad (13)$$

And from the second equation of (11), we know

$$\omega \tau < \frac{(2n+1)\pi}{2} \quad (14)$$

Therefore, Eq.(6) may have roots with positive real part except for $\tau = \tau_0$. Finally, we now satisfy the transversality condition of the Hopf spectrum, i.e.,

$$\text{Re}\left(\frac{d\lambda}{d\tau}\right)_{\tau=\tau_0} \neq 0$$

Differentiating Eq.(6) with respect to $\tau$ and applying the implicit function theorem, we have

$$\frac{d\lambda}{d\tau} = -\frac{b_2 \lambda e^{-\lambda\tau}}{1 + b_2 \tau e^{-\lambda\tau}}$$

Let $\lambda = \alpha + i\omega$, thus we have

$$\frac{d\lambda}{d\tau} = -\frac{b_2(\alpha + i\omega)e^{-\alpha\tau}(\cos(\omega\tau) - i\sin(\omega\tau))}{1 + b_2 \tau e^{-\alpha\tau}(\cos(\omega\tau) - i\sin(\omega\tau))}$$

From this we can obtain

$$\text{Re}\left(\frac{d\lambda}{d\tau}\right) = -\frac{b_2 e^{-\alpha\tau}(\alpha\cos(\omega\tau) + \omega\sin(\omega\tau) + b_2\alpha\tau e^{-\alpha\tau})}{(1 + b_2\tau e^{-\alpha\tau}\cos(\omega\tau))^2 + (b_2\tau e^{-\alpha\tau}\sin(\omega\tau))^2}$$

When $\tau = \tau_0$, $\sin\omega_0\tau_0 = -\omega_0/b_2$. It is obvious that

$$\text{Re}\left(\frac{d\lambda}{d\tau}\right)\bigg|_{\tau=\tau_0} = \frac{\omega_0^2}{(1 - h\tau_0)^2 + (\omega_0\tau_0)^2} > 0$$

Similarly, we can get

$$\text{Im}\left(\frac{d\lambda}{d\tau}\right) = -\frac{b_2 e^{-\alpha\tau}(\omega\cos(\omega\tau) - \alpha\sin(\omega\tau) + b_2\omega\tau e^{-\alpha\tau})}{(1 + b_2\tau e^{-\alpha\tau}\cos(\omega\tau))^2 + (b_2\tau e^{-\alpha\tau}\sin(\omega\tau))^2}$$

Thus

$$\text{Im}\left(\frac{d\lambda}{d\tau}\right)\bigg|_{\tau=\tau_0} = \frac{\omega_0\left(h - b_2^2\tau\right)}{(1 - h\tau_0)^2 + (\omega_0\tau_0)^2}$$

Based on above analysis, we can get the following theorem by applying the Hopf bifurcation theorem for delay differential equations [31].

**Theorem 1.** When the communication delay $\tau$ is smaller than the critical value $\tau_0$ described in (9), the equilibrium point $p^*$ of controlled system (2) is asymptotically stable. When the delay $\tau$ passes through $\tau_0$, there is a Hopf bifurcation of controlled system (2) at its equilibrium point $p^*$.

When $h = 0$, the controlled system (2) becomes original uncontrolled system (1). Then we get the following corollary.

**Corollary 1.** When the communication delay $\tau$ is smaller than the critical value $\tau_0 = -\pi/(2b)$, the equilibrium point $p^*$ of system (1) is asymptotically stable. When the delay $\tau$ passes through $\tau_0$, there is a Hopf bifurcation of system (1) at its equilibrium point $p^*$.

## 3. Stability and Direction of bifurcating periodic solutions

In this section, by using the normal form theory and the center manifold theorem, we determinate the direction of the bifurcation and the stability of bifurcating periodic solutions of controlled system (2) as it transits from stability to instability via a Hopf bifurcation.

For convenience, let $\tau = \tau_0 + \mu$. Thus $\mu = 0$ is the Hopf bifurcation value for Eq. (4). Let

$$L_\mu \phi = h\phi(0) + b_2\phi(-\tau)$$

and

$$F(\phi, \mu) = b_4\phi(0)\phi(-\tau) + b_5\phi^2(-\tau) + b_8\phi(0)\phi^2(-\tau) + b_9\phi^3(-\tau) + O(|\phi|^4)$$

Therefore

$$\dot{u}(t) = L_\mu u_t + F(u_t, \mu) \tag{15}$$

By the Riese representation theorem, there is a function of bounded variation $\eta(\theta, \mu)$ with $\theta \in [-\tau, 0]$ such that

$$L_\mu \phi = \int_{-\tau}^{0} d\eta(\theta, \phi)\phi(\theta)$$

which can be satisfied by choosing

$$d\eta(\theta, \phi) = h\delta(0) + b_2\delta(-\tau)$$

where $\delta$ is the Dirac delta function.

For $\phi \in C^1([-\tau, 0], \mathbb{R})$, define

$$A(\mu)\phi = \begin{cases} \dfrac{d\phi}{d\theta}, & \theta \in [-\tau, 0) \\ \displaystyle\int_{-\tau}^{0} d\eta(\theta, \mu)\phi(\theta), & \theta = 0 \end{cases} \tag{16}$$

and

$$R(\mu)\phi = \begin{cases} 0, & \theta \in [-\tau, 0) \\ F(\mu, \phi), & \theta = 0 \end{cases} \tag{17}$$

Thus Eq.(4) can be rewritten as

$$\dot{u}(t) = A(\mu)u_t + R(\mu)u_t \tag{18}$$

where $u_t = u(t + \theta)$ for $\theta \in [-\tau, 0]$.

For $\psi \in C^1([0,\tau],\mathbb{R})$, the adjoint operator $A^*$ of $A$ is defined as

$$A^*(\mu)\psi(s) = \begin{cases} \dfrac{d\phi(s)}{d\theta}, & s \in (0,\tau] \\ \int_{-\tau}^{0} d\eta(s,\mu)\psi(-s), & s = 0 \end{cases}$$

For $\phi \in C^1([-\tau,0],\mathbb{R})$ and $\psi \in C^1([0,\tau],\mathbb{R})$, we define a bilinear inner product by

$$\langle \psi, \varphi \rangle = \bar{\psi}(0)\varphi(0) - \int_{\theta=-\tau}^{0} \int_{s=0}^{\theta} \bar{\psi}(s-\theta) d\eta(\theta) \varphi(s) ds \qquad (19)$$

where $d\eta(\theta) = d\eta(\theta,0)$.

In order to determine the Poincare normal form of the operator $A(0)$, we need to calculate the eigenvector $q(\theta)$ of $A(0)$ corresponding to the eigenvalue $i\omega_0$ and the eigenvector $q^*(s)$ of $A^*(0)$ corresponding to the eigenvalue $-i\omega_0$. We can easily verify that

$$q(\theta) = \exp(i\omega_0 \theta), \ \theta \in [-\tau, 0)$$

and

$$q(s) = B\exp(-i\omega_0 s), \ s \in (0,\tau]$$

From Eq.(19), we have

$$\begin{aligned} \langle q^*, q \rangle &= \bar{q}^*(0)q(0) - \int_{\theta=-\tau}^{0} \int_{s=0}^{\theta} \bar{q}^*(s-\theta) d\eta(\theta) q(s) ds \\ &= \bar{B} - \int_{\theta=-\tau}^{0} \int_{s=0}^{\theta} \bar{B}\exp(-i\omega_0(s-\theta)) d\eta(\theta) \exp(i\omega_0 \tau) ds \\ &= \bar{B} - \bar{B}\int_{\theta=-\tau}^{0} \theta \exp(-i\omega_0 \tau) d\eta(\theta) \\ &= \bar{B}(1 + b_2 \tau \exp(-i\omega_0 \tau)) \end{aligned}$$

Thus we choose

$$B = \frac{1}{1 + b_2 \tau \exp(i\omega_0 \tau)}$$

such that $\langle q^*, q \rangle = 1$.

Similarly, we now prove that $\langle q^*, \bar{q} \rangle = 0$. Also using Eq.(19), we get

$$\langle q^*, \overline{q} \rangle = \overline{q}^*(0)\overline{q}(0) - \int_{\theta=-\tau}^{0}\int_{s=0}^{\theta}\overline{q}^*(s-\theta)d\eta(\theta)\overline{q}(s)ds$$

$$= \overline{B} - \int_{\theta=-\tau}^{0}\int_{s=0}^{\theta}\overline{B}\exp(-i\omega_0(s-\theta))d\eta(\theta)\exp(-i\omega_0\tau)ds$$

$$= \overline{B} + \frac{\overline{B}}{i2\omega_0}\int_{\theta=-\tau}^{0}(\exp(-i\omega_0\tau)-\exp(i\omega_0\tau))d\eta(\theta)$$

$$= \overline{B}\left[1 + \frac{b_2(\exp(-i\omega_0\tau)-\exp(i\omega_0\tau))}{i2\omega_0}\right]$$

Since $A(0)q(0) = i\omega_0 q(0)$ and $A^*(0)q(0) = -i\omega_0 q^*(0)$, we get

$$h + b_2\exp(-i\omega_0\tau) = i\omega_0 \text{ and } h + b_2\exp(i\omega_0\tau) = -i\omega_0$$

Hence

$$b_2(\exp(-i\omega_0\tau)-\exp(i\omega_0\tau)) = -i2\omega_0$$

Therefore $\langle q^*, \overline{q} \rangle = 0$.

Let $z(t) = \langle q^*, u_t \rangle$ and define

$$W(t,\theta) = u_t - zq - \overline{z}\overline{q} = u_t - 2\text{Re}\{z(t)q(\theta)\}$$

Then, on the manifold $C_0$, $W(t,\theta) = W(z(t), \overline{z}(t), \theta)$ where

$$W(z,\overline{z},\theta) = W_{20}(\theta)\frac{z^2}{2} + W_{11}(\theta)z\overline{z} + W_{02}(\theta)\frac{\overline{z}^2}{2} + \cdots \quad (20)$$

Here $z$ and $\overline{z}$ are local coordinates for $C_0$ in $C$ in the directions of $q$ and $\overline{q}^*$, respectively. Note that $W$ is real if $u_t$ is real; we deal with real solutions only. At $\mu = 0$

$$\begin{aligned}\dot{z}(t) &= \langle q^*, \dot{u}_t \rangle \\ &= \langle q^*, Au_t + Ru_t \rangle \\ &= i\omega_0 z(t) + \overline{q}^*(0)F_0(z,\overline{z})\end{aligned} \quad (21)$$

which can be written as in abbreviated form

$$\dot{z}(t) = i\omega_0 z(t) + g(z,\overline{z}) \quad (22)$$

The next objective is to expand $g$ in powers of $z$ and $\overline{z}$.

$$g(z,\overline{z}) = \overline{q}^*(0)F_0(z,\overline{z})$$
$$= g_{20}\frac{z^2}{2} + g_{11}z\overline{z} + g_{02}\frac{\overline{z}^2}{2} + g_{21}\frac{z^2\overline{z}}{2} + \cdots \quad (23)$$

Following the algorithms giving in [31], we write

$$\dot{W} = \dot{u}_t - \dot{z}q - \dot{\overline{z}}\,\overline{q} \tag{24}$$

Using (18) and (22), we obtain

$$\dot{W} = \begin{cases} AW - 2\operatorname{Re}\{\overline{q}*(0)F_0 q(\theta)\}, & \theta \in [-\tau, 0) \\ AW - 2\operatorname{Re}\{\overline{q}*(0)F_0 q(0)\} + F_0, & \theta = 0 \end{cases}$$

which is rewritten as

$$\dot{W} = AW + H(z, \overline{z}, \theta) \tag{25}$$

where

$$H(z, \overline{z}, \theta) = H_{20}(\theta)\frac{z^2}{2} + H_{11}(\theta)z\overline{z} + H_{02}(\theta)\frac{\overline{z}^2}{2} + \cdots \tag{26}$$

On the other hand, on $C_0$

$$\dot{W} = W_z \dot{z} + W_{\dot{z}}\,\dot{\overline{z}} \tag{27}$$

Using (20), (22) to replace $W_z$ and $\dot{z}$ and their conjugates by their power series expansions, we get a second expression for $\dot{W}$

$$\dot{W} = i\omega_0 W_{20}(\theta)z^2 - i\omega_0 W_{02}(\theta)\overline{z}^2 + \cdots \tag{28}$$

Comparing the coefficients of the above equation with those of (25), we get

$$\begin{cases} (A - i2\omega_0)W_{20}(\theta) = -H_{20}(\theta) \\ W_{11}(\theta) = -H_{11}(\theta) \\ (A + i2\omega_0)W_{02}(\theta) = -H_{02}(\theta) \end{cases} \tag{29}$$

Observing

$$u_t(\theta) = W(z, \overline{z}, \theta) + zq(\theta) + \overline{z} \cdot \overline{q}(\theta)$$
$$= W_{20}(\theta) \cdot \frac{z^2}{2} + W_{11}(\theta)z\overline{z} + W_{02}(\theta) \cdot \frac{\overline{z}^2}{2}$$
$$+ z\exp(i\omega_0 \theta) + \overline{z}\exp(-i\omega_0 \theta) + \cdots$$

from which we obtain $u_t(0)$ and $u_t(-\tau)$

$$u_t(0) = W(z, \overline{z}, 0) + z + \overline{z}$$

$$u_t(-\tau) = W(z, \overline{z}, -\tau) + z\exp(-i\omega_0 \tau) + \overline{z}\exp(i\omega_0 \tau)$$

As we only need the coefficients of $z^2$, $z\overline{z}$, $\overline{z}^2$ and $z^2\overline{z}$, we keep these relevant terms in the following expansions

$$u_t(0)u_t(-\tau) = z^2 \exp(-i\omega_0\tau) + z\bar{z}(\exp(i\omega_0\tau) + \exp(-i\omega_0\tau)) + \bar{z}^2 \exp(i\omega_0\tau)$$
$$+ z^2\bar{z}\left[W_{11}(0)\exp(-i\omega_0\tau) + \frac{W_{20}(0)}{2}\exp(i\omega_0\tau) + W_{11}(-\tau) + \frac{W_{20}(-\tau)}{2}\right] + \cdots$$

$$u_t^2(-\tau) = z^2 \exp(-i2\omega_0\tau) + \bar{z}^2 \exp(i2\omega_0\tau) + 2z\bar{z}$$
$$+ z^2\bar{z}\left[2\exp(-i\omega_0\tau)W_{11}(-\tau) + \exp(i\omega_0\tau)W_{20}(-\tau)\right] + \cdots$$

$$u_t(0)u_t^2(-\tau) = z^2\bar{z}(\exp(-i2\omega_0\tau) + 2) + \cdots$$

$$u_t^3(-\tau) = 3z^2\bar{z}\exp(-i\omega_0\tau) + \cdots$$

Therefore we have

$$g(z,\bar{z}) = z^2 \bar{B}\left(b_4 \exp(-i\omega_0\tau) + b_5 \exp(-i2\omega_0\tau)\right)$$
$$+ z\bar{z}\bar{B}\left[b_4\left(\exp(i\omega_0\tau) + \exp(-i\omega_0\tau)\right) + 2b_5\right] + \bar{z}^2\bar{B}\left(b_4 \exp(i\omega_0\tau) + b_5 \exp(i2\omega_0\tau)\right)$$
$$+ z^2\bar{z}\bar{B}\Big[b_4 W_{11}(0)\exp(-i\omega_0\tau) + \frac{W_{20}(0)}{2}\exp(i\omega_0\tau) + W_{11}(-\tau) + \frac{W_{20}(-\tau)}{2}$$
$$+ b_5\left(2W_{11}(-\tau)\exp(-i\omega_0\tau) + W_{20}(-\tau)\exp(i\omega_0\tau)\right) + b_8(\exp(-i2\omega_0\tau) + 2) + 3b_9\exp(-i\omega_0\tau)\Big]$$

Comparing above coefficients with those in (23), we get

$$g_{20} = 2\bar{B}\left(b_4 \exp(-i\omega_0\tau) + b_5 \exp(-i2\omega_0\tau)\right) \tag{30}$$

$$g_{11} = \bar{B}\left[b_4\left(\exp(i\omega_0\tau) + \exp(-i\omega_0\tau)\right) + 2b_5\right] \tag{31}$$

$$g_{02} = 2\bar{B}\left(b_4 \exp(i\omega_0\tau) + b_5 \exp(i2\omega_0\tau)\right) \tag{32}$$

$$g_{21} = 2\bar{B}\Big[b_4 W_{11}(0)\exp(-i\omega_0\tau) + \frac{W_{20}(0)}{2}\exp(i\omega_0\tau) + W_{11}(-\tau) + \frac{W_{20}(-\tau)}{2}$$
$$+ b_5\left(2W_{11}(-\tau)\exp(-i\omega_0\tau) + W_{20}(-\tau)\exp(i\omega_0\tau)\right) + b_8(\exp(-i2\omega_0\tau) + 2) + 3b_9\exp(-i\omega_0\tau)\Big] \tag{33}$$

We still need to compute $W_{20}(0)$, $W_{20}(-\tau)$, $W_{11}(0)$ and $W_{11}(-\tau)$ for the expression of $g_{21}$.

For $\theta \in [-\tau, 0)$,

$$H(z,\bar{z},\theta) = -2\operatorname{Re}\{\bar{q}^*(0)F_0 q(\theta)\}$$
$$= -2\operatorname{Re}\{g(z,\bar{z})q(\theta)\}$$
$$= -g(z,\bar{z})q(\theta) - \bar{g}(z,\bar{z})\bar{q}(\theta)$$
$$= -\left(g_{20}\frac{z^2}{2} + g_{11}z\bar{z} + g_{02}\frac{\bar{z}^2}{2} + \cdots\right)q(\theta)$$
$$-\left(\bar{g}_{20}\frac{\bar{z}^2}{2} + \bar{g}_{11}z\bar{z} + \bar{g}_{02}\frac{z^2}{2} + \cdots\right)\bar{q}(\theta)$$

Comparing the coefficients of above equation with those of Eq.(26), we obtain

$$\begin{cases} H_{20}(\theta) = -g_{20}q(\theta) - \overline{g}_{02}\overline{q}(\theta) \\ H_{11}(\theta) = -g_{11}q(\theta) - \overline{g}_{11}\overline{q}(\theta) \end{cases}$$

From (16) and (29), we get

$$\dot{W}_{20}(\theta) = i2\omega_0 W_{20}(\theta) + g_{20}q(\theta) + \overline{g}_{02}\overline{q}(\theta) \tag{34}$$

and

$$\dot{W}_{11}(\theta) = g_{11}q(\theta) + \overline{g}_{11}\overline{q}(\theta) \tag{35}$$

Solving Eq.(34) and (35), we have

$$W_{20}(\theta) = -\frac{g_{20}}{i\omega_0}q(0)\exp(i\omega_0\theta) - \frac{\overline{g}_{02}}{i3\omega_0}\overline{q}(0)\exp(-i\omega_0\theta) + E_1\exp(i2\omega_0\theta) \tag{36}$$

and

$$W_{11}(\theta) = \frac{g_{11}}{i\omega_0}q(0)\exp(i\omega_0\theta) - \frac{\overline{g}_{11}}{i\omega_0}\overline{q}(0)\exp(-i\omega_0\theta) + E_2 \tag{37}$$

where $E_1$ and $E_2$ are both constants, and can be determined by setting $\theta = 0$ in $H(z,\overline{z},\theta)$. It is evident that

$$H(z,\overline{z},0) = -2\operatorname{Re}\{\overline{q}^*(0)F_0 q(0)\} + F_0$$

Thus

$$H_{20}(0) = -g_{20}q(0) - \overline{g}_{20}\overline{q}(0) + 2\left(b_4\exp(-i\omega_0\tau) + b_5\exp(-i2\omega_0\tau)\right) \tag{38}$$

$$H_{11}(0) = -g_{11}q(0) - \overline{g}_{11}\overline{q}(0) + \left[b_4\left(\exp(i\omega_0\tau) + \exp(-i\omega_0\tau)\right) + 2b_5\right] \tag{39}$$

From (29) and recall that

$$AW_{20}(0) = hW_{20}(0) + b_2 W_{20}(-\tau)$$

$$AW_{11}(0) = hW_{11}(0) + b_2 W_{11}(-\tau)$$

we get

$$\begin{aligned} &hW_{20}(0) + b_2 W_{20}(-\tau) - i2\omega_0 W_{20}(0) \\ &= g_{20}q(0) + \overline{g}_{20}\overline{q}(0) - 2\left(b_4\exp(-i\omega_0\tau) + b_5\exp(-i2\omega_0\tau)\right) \end{aligned} \tag{40}$$

and

$$\begin{aligned} &hW_{11}(0) + b_2 W_{11}(-\tau) \\ &= g_{11}q(0) + \overline{g}_{11}\overline{q}(0) - \left[b_4\left(\exp(i\omega_0\tau) + \exp(-i\omega_0\tau)\right) + 2b_5\right] \end{aligned} \tag{41}$$

Substituting (36) into (40), we obtain

$$E_1 = \frac{\Phi_1}{h + b_2\exp(-i2\omega_0\tau) - i2\omega_0}$$

where

$$\Phi_1 = (h - i2\omega_0)\left(\frac{g_{20}}{i\omega_0} + \frac{\bar{g}_{02}}{i3\omega_0}\right) + b_2\left(\frac{g_{20}}{i\omega_0}\exp(-i\omega_0\tau) + \frac{\bar{g}_{02}}{i3\omega_0}\exp(i\omega_0\tau)\right)$$
$$+ g_{20} + \bar{g}_{02} - 2(b_4\exp(-i\omega_0\tau) + b_5\exp(-i2\omega_0\tau))$$

Similarly, substituting (37) into (41), we get

$$E_2 = \frac{\Phi_2}{h + b_2}$$

where

$$\Phi_2 = -h\left(\frac{g_{11}}{i\omega_0} - \frac{\bar{g}_{11}}{i\omega_0}\right) - b_2\left(\frac{g_{11}}{i\omega_0}\exp(i\omega_0\tau) - \frac{\bar{g}_{11}}{i\omega_0}\exp(-i\omega_0\tau)\right)$$
$$+ g_{11} + \bar{g}_{11} - \left[b_4(\exp(i\omega_0\tau) + \exp(-i\omega_0\tau)) + 2b_5\right]$$

Therefore we have formulas to compute the following parameters:

$$\begin{aligned}
C_1(0) &= \frac{i}{2\omega_0}\left(g_{20}g_{11} - 2|g_{11}|^2 - \frac{1}{3}|g_{02}|^2\right) + \frac{g_{21}}{2} \\
\mu_2 &= -\frac{\operatorname{Re}\{C_1(0)\}}{\operatorname{Re}\lambda'(0)} \\
T_2 &= -\frac{\operatorname{Im}\{C_1(0)\} + \mu_2\operatorname{Im}\lambda'(0)}{\omega_0} \\
\beta_2 &= 2\operatorname{Re}\{C_1(0)\}
\end{aligned} \quad (42)$$

where $C_1(0)$ is the Lyapunov coefficient. Now we give the main results of this section.

**Theorem 2.** For controlled system (2), when $\tau = \tau_0$, the direction and stability of periodic solutions of the Hopf bifurcation is determinated by the formulas (42) and the following results hold:

(1). $\mu_2$ determines the direction of the Hopf bifurcation. If $\mu_2 > 0 (< 0)$, the Hopf bifurcation is supercritical (subcritical) and the bifurcating periodic solutions exist for $\tau > \tau_0 (\tau < \tau_0)$.

(2). $\beta_2$ determines the stability of the bifurcating periodic solution. If $\beta_2 < 0 (> 0)$, the bifurcating periodic solutions are stable (unstable).

(3). $T_2$ determines the period of the bifurcating periodic solution. If $T_2 > 0 (< 0)$, the period increases (decreases).

4. **Numerical Simulations**

In this section, some numerical examples are given to verify the results obtained in the previous section. We consider a fair dual which give a proportionally fair resource allocation, i.e., $x(t) = 1/p(t)$ [23].

Let the link capacity is $1.25\ Mbps$ and the time unit is $40\ ms$. If the packet sizes are 1000 bytes each, the link capacity can be expressed as $c = 50$ packets per time unit. In addition, let the gain parameter $k = 0.1$.

We first choose $h = 0$ which represents the system without control. By direct calculation we get

$$p^* = 0.02,\ \omega_0 = 0.5,\ \tau_0 = 3.1416$$

$$\mu_2 = 5259.2,\ T_2 = 2125,\ \beta_2 = -758.38$$

The dynamic behavior of the uncontrolled model (1) is illustrated in Fig.1-Fig.3. From Corollary 1, it is obvious that when $\tau < \tau_0$, trajectories converge to the equilibrium point (see Fig.1), while as $\tau$ is increased to pass $\tau_0$, $p^*$ loses its stability and a Hopf bifurcation occurs (see Fig.2 and Fig.3). Since $\beta_2 < 0$, the periodic orbits are stable. As $\mu_2 > 0$, the Hopf bifurcation is supercritical and the bifurcation periodic solutions exist when $\tau > \tau_0$. The period of the periodic solutions increases as $\tau$ increases due to $T_2 > 0$ ( compare Fig.2 and Fig.3).

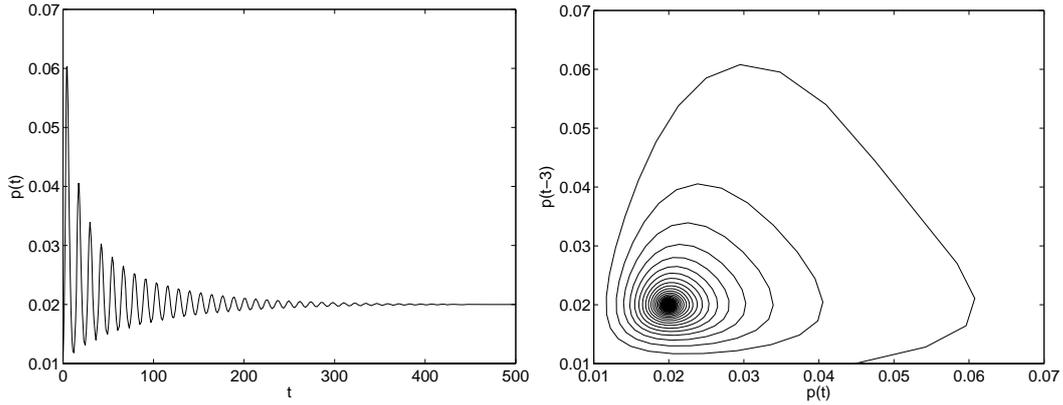

Fig.1 Waveform plot and phase portrait of uncontrolled system (1) with $\tau = 3$

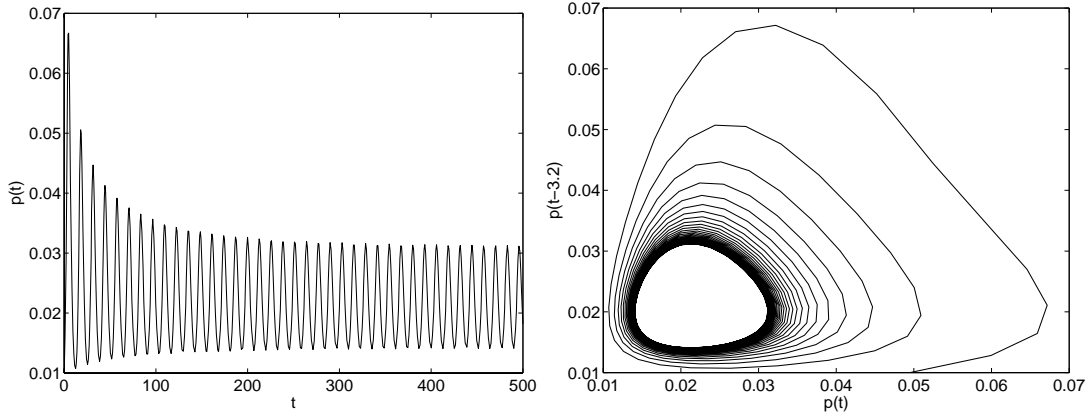

Fig.2 Waveform plot and phase portrait of uncontrolled system (1) with $\tau = 3.2$

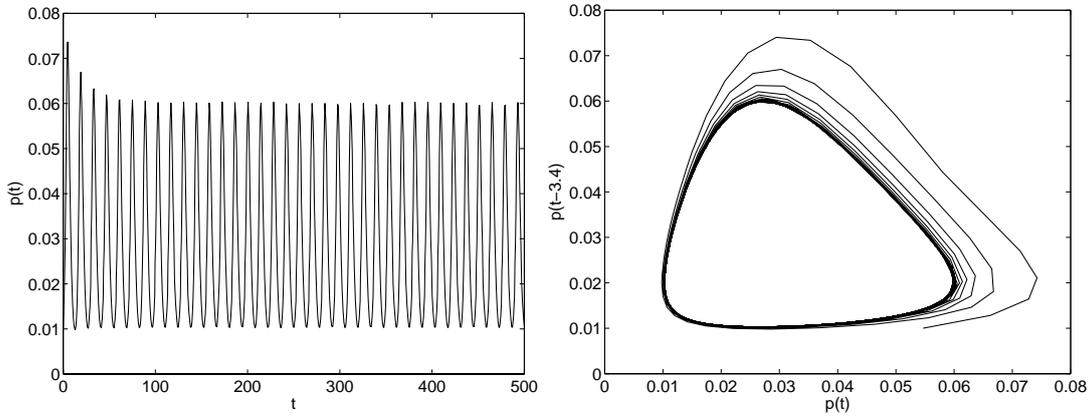

Fig.3 Waveform plot and phase portrait of uncontrolled system (1) with $\tau = 3.4$

Then we consider the problem of controlling the Hopf bifurcation in system (1). From (10), we see that the feedback gain is in the range of $[-0.25, 0)$. Therefore, we can delay the onset of Hopf bifurcation by choosing an appropriate value of $h$. For example, by choosing $h = -0.1$, we obtain

$$p^* = 0.02, \quad \omega_0 = 0.3873, \quad \tau_0 = 4.7082$$

$$\mu_2 = 27606, \; T_2 = 5572.9, \; \beta_2 = -1508.9$$

Note that the controlled system (2) has the same equilibrium point as that of the original system (1), but the critical value $\tau_0$ increases from $3.1416$ to $4.7082$, implying that the onset of Hopf bifurcation is delayed. It is seen from Fig.4 that when $\tau = 3.4$ (the same value used in Fig.3), instead of having a limit cycle, the controlled system (2) converges to the equilibrium point $p^*$. When $\tau$ passes the critical value $\tau_0 = 4.7082$, a Hopf bifurcation occurs (see Fig.5 and Fig.6).

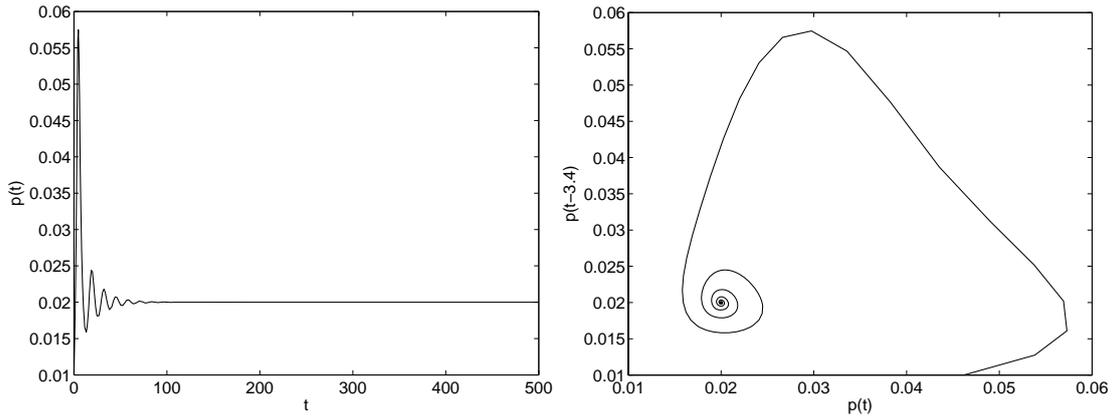

Fig.4 Waveform plot and phase portrait of controlled system (2) with $\tau = 3.4$ and $h = -0.1$

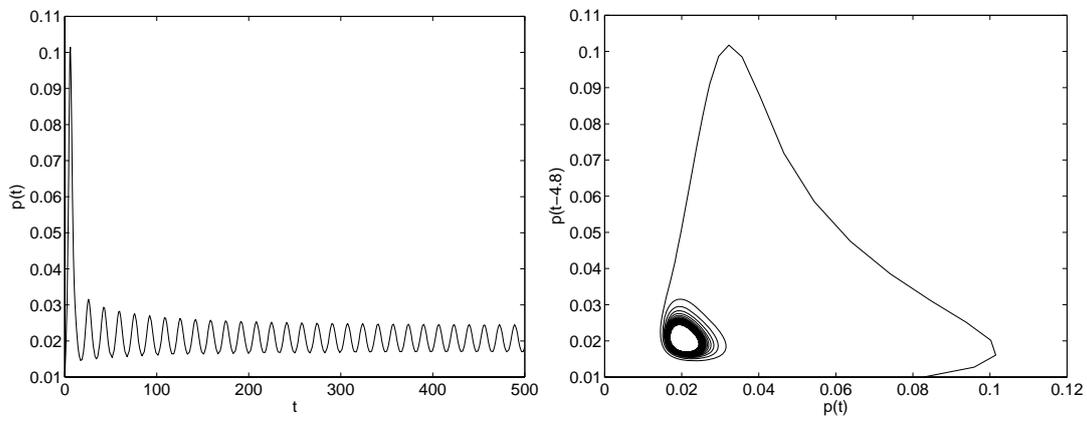

Fig.5 Waveform plot and phase portrait of controlled system (2) with $\tau = 4.8$ and $h = -0.1$

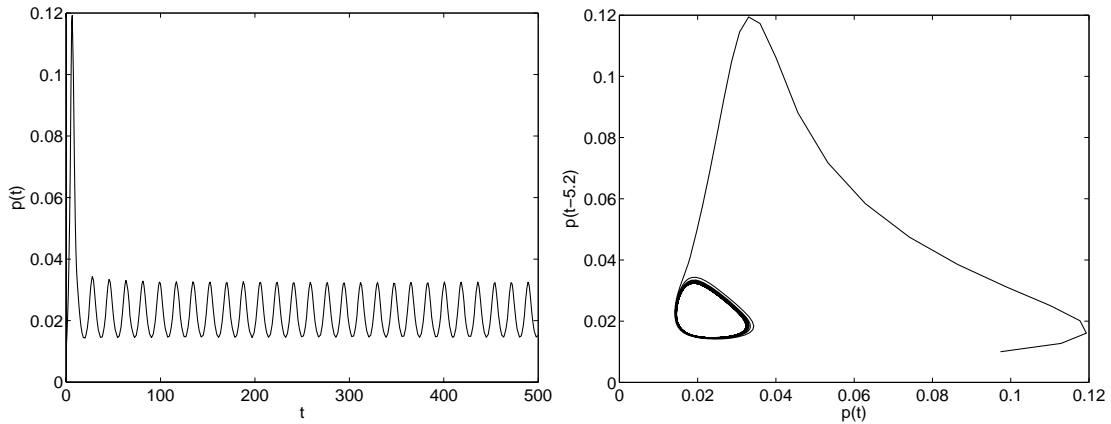

Fig.6 Waveform plot and phase portrait of controlled system (2) with $\tau = 5.2$ and $h = -0.1$

The relationship between the critical value $\tau_0$ and the control feedback gain $h$ is shown in Fig.7. From this figure we know that when decreasing $h$, the critical value increases, and then we can get a larger stability range of the system. For example, by choosing $h = -0.15$, we get that the critical value $\tau_0 = 6.3679$. In detail, if we set $\tau = 5.2$ (the same value used in Fig.6), which is less than the critical value $\tau_0$, the controlled system (2) converges to the equilibrium point, as shown in Fig.8.

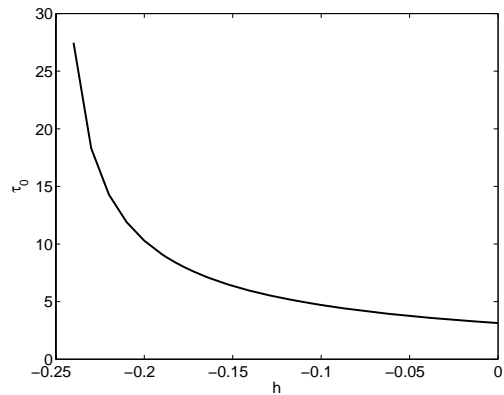

Fig.7 The fluctuation of $\tau_0$ depending on $h$

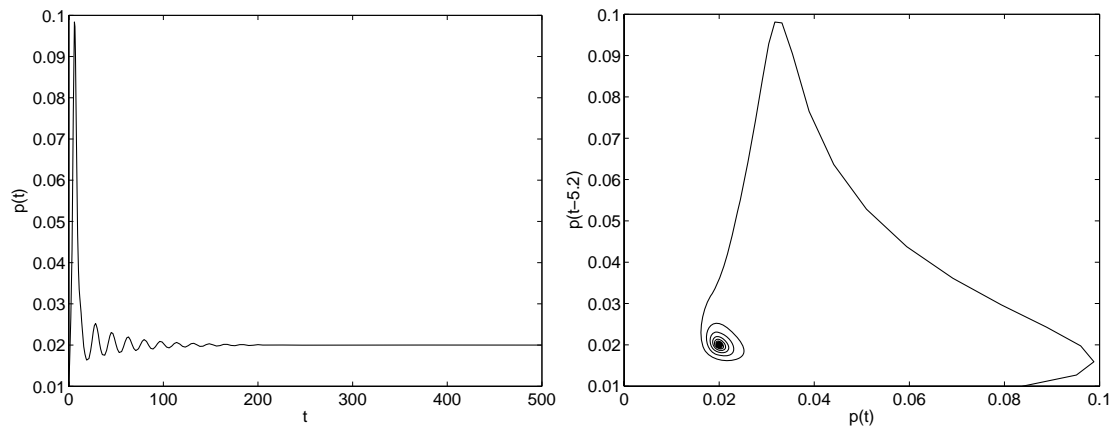

Fig.8 Waveform plot and phase portrait of controlled system (2) with $\tau = 5.2$ and $h = -0.15$

## 5. Conclusions

In this paper, a time-delayed feedback controller has been proposed for a fair dual model of Internet congestion control system. By choosing communication delay as bifurcation parameter, it has been demonstrated that both the uncontrolled system and the controlled system lose stability and a Hopf bifurcation occurs when the delay passes a critical value. However, from theoretical analysis and numerical simulations, it has been shown that the critical value of the system with control is larger than that of the system without control, which indicates that the onset of undesired Hopf bifurcation has been delayed. Meanwhile, the direction of the Hopf bifurcation and the stability of the bifurcating periodic solutions have also been determinated by using the center manifold theorem and the normal form theory.

**Acknowledgments**

This work was supported by the National Natural Science Foundation of China with the grant numbers 70571017.